\documentclass[conference]{IEEEtran}
\IEEEoverridecommandlockouts
\usepackage{cite}
\usepackage{amsmath,amssymb,amsfonts}
\usepackage{graphicx}
\usepackage{textcomp}
\usepackage{xcolor}

\usepackage{hyperref}
\usepackage{multirow}
\usepackage{caption}
\usepackage{subcaption}
\usepackage{tabularx}
\usepackage{mathtools}
\usepackage{color}
\usepackage{multicol}
\usepackage{flushend}
\usepackage{listings} 
\usepackage{algorithm}
\usepackage[noend]{algpseudocode}
\usepackage{booktabs}
\usepackage{microtype}
\usepackage{eso-pic}

\makeatletter
\def\ps@IEEEtitlepagestyle{%
	\def\@oddfoot{\mycopyrightnotice}%
	\def\@oddhead{\hbox{}\@IEEEheaderstyle\leftmark\hfil\thepage}\relax
	\def\@evenhead{\@IEEEheaderstyle\thepage\hfil\leftmark\hbox{}}\relax
	\def\@evenfoot{}%
}
\def\mycopyrightnotice{%
	\begin{minipage}{\textwidth}
		\scriptsize
		\copyright~2024 IEEE. DOI: 10.1109/ICECS61496.2024.10848613.
		Personal use of this material is permitted. 
		Permission from IEEE must be obtained for all other uses,
		in any current or future media, including reprinting/republishing this material
		for advertising or promotional purposes, creating new collective works,
		for resale or redistribution to servers or lists, or reuse of any copyrighted
		component of this work in other works by sending a request to pubs-permissions@ieee.org.
	\end{minipage}
}
\makeatother

\def\BibTeX{{\rm B\kern-.05em{\sc i\kern-.025em b}\kern-.08em
    T\kern-.1667em\lower.7ex\hbox{E}\kern-.125emX}}
\begin{document}

\title{Functional ISS-Driven Verification of\\Superscalar RISC-V Processors}

\author{
	\IEEEauthorblockN{Andrea Galimberti}
	\IEEEauthorblockA{\textit{DEIB}\\
			\textit{Politecnico di Milano}\\
			Milano, Italy\\
			andrea.galimberti@polimi.it}
	\and
	\IEEEauthorblockN{Marco Vitali}
	\IEEEauthorblockA{\textit{DEIB}\\
			\textit{Politecnico di Milano}\\
			Milano, Italy\\
			marco9.vitali@mail.polimi.it}
	\and
	\IEEEauthorblockN{Sebastiano Vittoria}
	\IEEEauthorblockA{\textit{DEIB}\\
		\textit{Politecnico di Milano}\\
		Milano, Italy\\
		sebastiano.vittoria@mail.polimi.it}
	\and
	\IEEEauthorblockN{Davide Zoni}
	\IEEEauthorblockA{\textit{DEIB}\\
			\textit{Politecnico di Milano}\\
			Milano, Italy\\
			davide.zoni@polimi.it}
}

\maketitle

\IEEEpeerreviewmaketitle

\begin{abstract}
A time-efficient and comprehensive verification is a fundamental part of
the design process for modern computing platforms, and it becomes ever more
important and critical to optimize as the latter get ever more complex.
\emph{SupeRFIVe} is a methodology for the functional verification of
superscalar processors that leverages an instruction set simulator to
validate their correctness according to a simulation-based approach,
interfacing a testbench for the design under test with
the instruction set simulator by means of socket communication.
We demonstrate the effectiveness of the \emph{SupeRFIVe} methodology by
applying it to verify the functional correctness of a RISC-V dual-issue superscalar CPU,
leveraging the state-of-the-art RISC-V instruction set simulator \emph{Spike}
and executing a set of benchmark applications from the open literature.
\end{abstract}


\section{Introduction}
\label{sec:introduction}
The emergence of increasingly more computationally intensive applications
in the last decades has led to reinvigorated research in the field of
computer architecture, with the goal of designing computing platforms that
provided better performance~\cite{Fornaciari_2023DATE} and
higher energy and power efficiency~\cite{Zoni_2023CSUR}.
To this end, superscalar architectures enable exploiting instruction-level parallelism by
issuing multiple instructions per clock cycle and executing them on their various functional units,
thus improving the throughput of the CPU cores.

At the same time, RISC-V has emerged as the de-facto standard instruction set architecture (ISA)
for both academic and industrial research thanks to its open-source and royalty-free nature,
as well as to its modular architecture that makes it easily extendable by
system designers according to their specific requirements.
The open literature provides a variety of RISC-V cores,
ranging from scalar in-order 32-bit resource-constrained processors~\cite{Zoni_2021SUSCOM,Zoni_2022JSA}
up to higher-performance 64-bit superscalar in-order~\cite{Andersson_2020DSN-W} and
speculative out-of-order ones~\cite{Zhao_2020CARRV}.

While processors evolve towards more complex architectures such as
superscalar ones to deal with novel workloads and time-to-market
deadlines get increasingly tighter, the functional verification of
their hardware design surges to an ever more prominent role in 
identifying errors time-efficiently and as early as possible in
their design and manufacturing process.
The methodologies for hardware design verification from the open literature
include simulation-based and formal verification approaches.
However, on the one hand, solutions for the verification of RISC-V architectures
are not meant for superscalar processors:
\emph{RISC-V Formal Verification Framework}~\cite{riscv-formal_GitHub} is
a framework for the formal verification of RISC-V processors that is
limited to the integer extensions of the RISC-V ISA,
supporting indeed solely the RV32IMC and RV64IMC architectures,
\cite{Jimenez_2023DAT} targets RISC-V vector processors, and
\cite{Duran_2020ISCAS} proposes a verification methodology that combines
the formal verification of \cite{riscv-formal_GitHub} and the simulation of the processor under test,
with input programs obtained by a genetic algorithm and a functional ISA simulator as the golden model,
and that is only applied to a single-issue, in-order, 3-stage scalar core.
On the other hand, the various existing simulation-based~\cite{Taylor_1998DAC} and
formal~\cite{Nelson_1997DAC} verification approaches meant for superscalar processors
target instead older architectures rather than RISC-V.

This manuscript proposes therefore a novel methodology for
the functional verification of RISC-V superscalar processors
that leverages a fast instruction set simulator (ISS)
to check the correctness of the design under test (DUT).
Notably, cycle-accurate simulators such as \emph{gem5} would
instead be too slow for such purposes, whereas
the open literature delivers multiple ISS solutions for
the RISC-V ISA~\cite{Spike_GitHub,Petersen_2021WCAE}.

\paragraph*{Contributions}
This paper introduces a novel methodology for functional design verification,
\mbox{\emph{SupeRFIVe}}~(\emph{\underline{Supe}rscalar \mbox{\underline{R}ISC-V} \underline{F}unctional \underline{I}SS-driven \underline{Ve}rification}), outlining two main contributions to the literature:
\begin{enumerate}
	\item \emph{SupeRFIVe} leverages an ISS to validate the correctness of
	a superscalar processor in a time-efficient and comprehensive way
	by interfacing, through socket communication, the ISS with a testbench
	written in a hardware verification language (HVL);
	\item we evaluate the effectiveness of the \emph{SupeRFIVe} methodology,
	applying it to a SystemVerilog RISC-V dual-issue superscalar CPU,
	by employing the de-facto standard RISC-V ISS \emph{Spike}~\cite{Spike_GitHub}
	and executing a set of applications from a state-of-the-art benchmark suite~\cite{Gustafsson_WCET2010}
	to carry out the functional verification as well as collect
	performance statistics related to the superscalar nature of
	the CPU design under verification.
\end{enumerate}

\section{SupeRFIVe Methodology}
\label{sec:methodology}
\begin{figure}[t]
	\centering
	\includegraphics[width=\columnwidth]{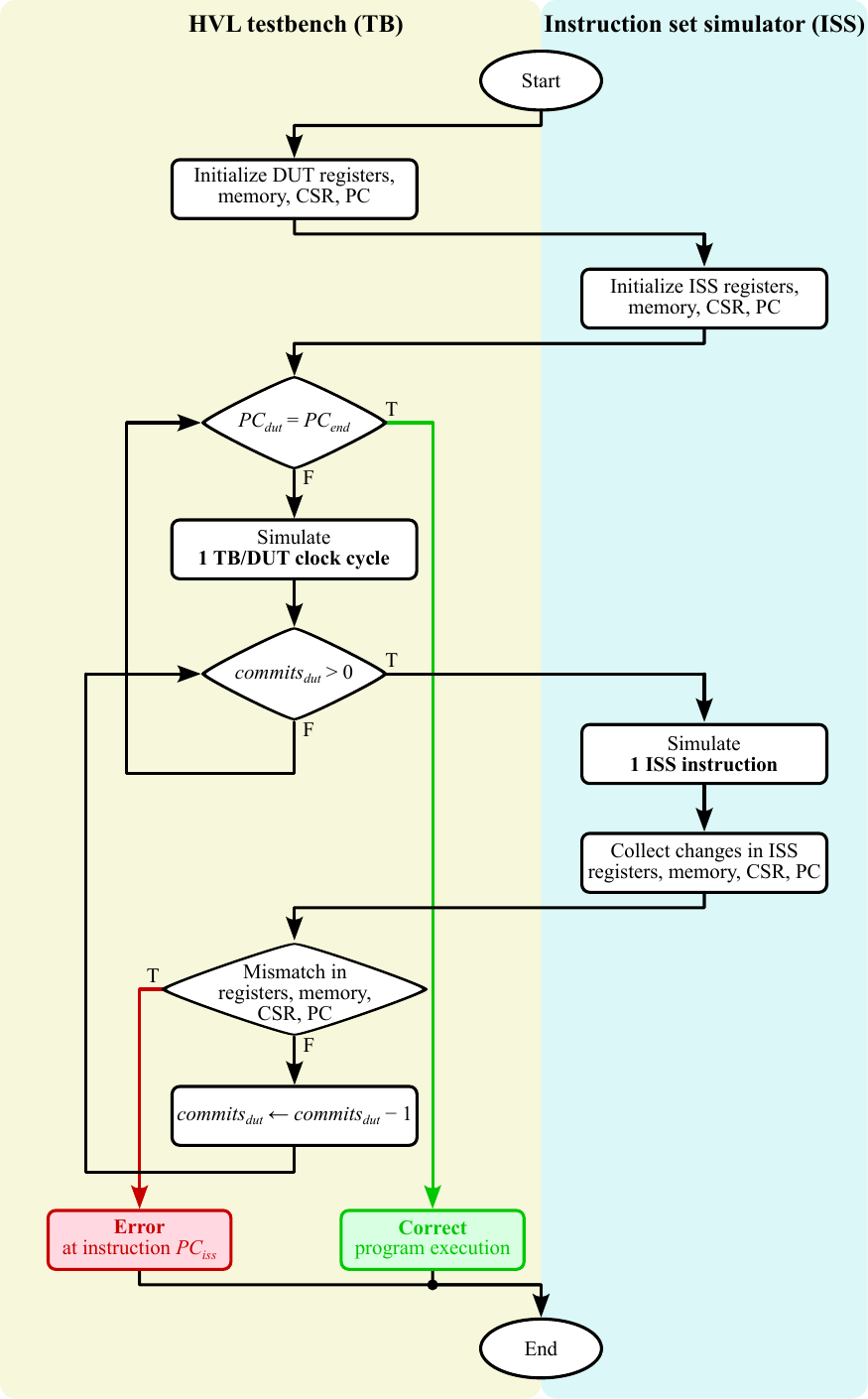}
	\caption{Flowchart of the \emph{SupeRFIVe} methodology.}
	\label{fig:verification}
\end{figure}

The proposed methodology performs the functional verification of
a superscalar processor (and, by extension, also of a scalar one)
by employing an ISS that supports the same ISA as the processor under test.
A socket-based communication enables interfacing and synchronizing
the ISS and an HVL testbench including the DUT processor,
so that the ISS and the DUT can compute the same instructions
from a target executable application in a lockstep fashion
in order to ease checking the results of instructions as
they get executed during the simulation.

\figurename~\ref{fig:verification} depicts the flow of the \emph{SupeRFIVe} methodology,
showing how the simulation of a target application's execution is carried out
in a synchronized manner in the instruction set simulator and
in the superscalar processor under verification to
verify the functional correctness of the latter.

Interleaving the simulation of the superscalar CPU under verification
with the execution of single instructions on the ISS side
allows immediately identifying errors in the DUT without executing the whole workload
and causes only a small overhead compared to a HVL testbench simulation that
does not interact with the ISS, in particular when moving from faster behavioral RTL simulations to
more computationally expensive post-synthesis and post-implementation ones,
in which the contribution of the ISS becomes negligible from an execution time standpoint.
Conversely, the adoption of an architectural simulator such as \emph{gem5}
rather than an ISS, i.e., a functional ISA simulator, would instead
drastically lengthen the verification time due to the cycle-level nature of
such simulators, albeit not providing any advantage with respect to
the functional verification of the target computing platform.

\subsubsection*{SupeRFIVe verification flow}
\begin{figure}[t]
	\centering
	\includegraphics[width=0.98\columnwidth]{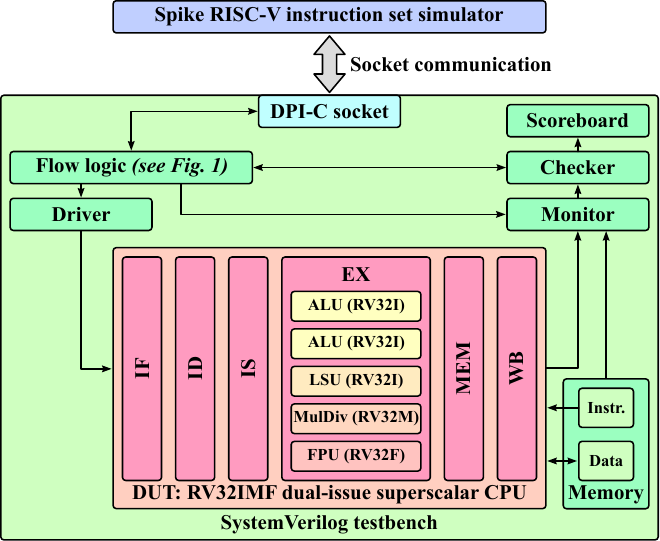}
	\caption{Verification infrastructure that implements
		the proposed \emph{SupeRFIVe} methodology for experimental evaluation purposes.}
	\label{fig:infrastructure}
\end{figure}

The HVL testbench~(TB) instantiates the DUT superscalar CPU and the main memory,
loads the executable file for the target application in the instruction memory, and
monitors the CPU registers and the associated memory during the execution of the application.
The same executable file is fed to the DUT, through the instruction memory, and to the ISS,
so that they execute the exact same application, enabling the continuous comparison
between the content of the CPU registers and memory in the TB and in the ISS to
verify the correct implementation of the target superscalar CPU.
The TB and the ISS communicate through a socket-based connection,
enabling their synchronization and the exchange of data related to
the instructions' results and the content of the memory and registers,
including the program counter~(PC) and the control and status registers~(CSRs).
An example of the verification infrastructure implementing
the \emph{SupeRFIVe} methodology is depicted in \figurename~\ref{fig:infrastructure}.

The process, depicted in \figurename~\ref{fig:verification},
to verify the correctness of the execution of a single application
on the target superscalar CPU involves, as the first step,
starting the simulation of the TB and the execution of the ISS,
opening sockets on both sides, and waiting until a socket communication is established.

On the TB side, a clock-cycle counter is reset to 0,
the program counter of the CPU (PC\textsubscript{dut}) is initialized to its starting value, and
the content of CPU registers and memory are also reset, while on the ISS side
the registers, memory, and PC value (PC\textsubscript{iss}) are correspondingly initialized.

The simulation of the testbench is carried out by checking, at each clock cycle,
whether one or more instructions have committed their results or not,
until the end of the executable file.
If no commits were performed in the TB simulation during
the current clock cycle, then the next clock cycle is simulated.
Otherwise, if one or more instructions have committed their results,
the ISS execution is advanced by a number of instructions that corresponds to
the number of commits on the TB side.
The result and the corresponding register or memory address of
each instruction executed on the ISS side are sent to the TB through the socket, and
the TB accordingly checks that its DUT register file and memory match the ISS ones.
Once a number of instructions have been executed on the ISS that is
equal to the number of concurrent commits in the DUT,
the TB simulation is finally restarted with the execution of the next clock cycle.

Mismatches in the PC values, CSRs, or in the register and memory contents
are signaled as errors and interrupt the verification process,
providing its user with informations related to the TB clock cycle,
TB and ISS PC values, and non-matching register or memory locations.
Conversely, if the application is executed correctly until its completion,
a set of statistics related to the instruction coverage and performance statistics,
e.g., execution time, CPI, and percentage of multiple commits,
is output by the verification framework.

\section{Experimental results}
\label{sec:expEval}
The proposed \emph{SupeRFIVe} methodology is evaluated by
applying it to a 32-bit RISC-V superscalar dual-issue CPU and
employing \emph{Spike} as the ISS that drives its verification in
a SystemVerilog testbench run in the \emph{AMD xsim} RTL simulator.

\subsection{Experimental setup}
\label{ssec:exp_setup}

\subsubsection*{Design under test and hardware setup}
The experimental evaluation targets, as the superscalar processor on
which to conduct the verification process,
a dual-issue superscalar CPU, described in SystemVerilog,
that implements a RV32IMF architecture, i.e., that supports
the baseline integer (I), integer multiplication and division (M),
and single-precision floating-point (F) extensions of the 32-bit RISC-V ISA.

The CPU has a pipelined architecture with six stages, namely
instruction fetch, instruction decode, issue,
execution, memory access, and write-back stages.
It features a 64-entry reorder buffer to commit up to two instruction results at a time and
a 16-entry issue queue to feed instructions and their operands to its functional units,
which include two arithmetic logic unit(s) (ALUs), a pipelined multiplication-division unit,
a load-store unit (LSU), and a pipelined single-precision floating-point unit.

Synthesis and implementation of the CPU were carried out in \emph{AMD Vivado 2023.1},
using the default synthesis and place-and-route optimization directives and
targeting an \emph{AMD Artix-7 75}~(\emph{xc7a75tftg256-1}) FPGA at a 77MHz clock frequency.
The target chip, from the mid-range cost-effective family of \emph{AMD} FPGAs and which
is commonly used both in the academia and in the industry, features
47200 look-up tables, 94400 flip-flops,
180 digital signal processing elements, and 105 36kb blocks of block RAM.

\subsubsection*{Software setup}
We employ version 1.1.0 of \emph{Spike}~\cite{Spike_GitHub} as the RISC-V ISS and
we implement a verification infrastructure according to the \emph{SupeRFIVe} methodology
by interfacing the SystemVerilog testbench for the CPU design under test with \emph{Spike}
and having them communicate through sockets that leverage,
on the testbench side, SystemVerilog's native DPI-C support.
The verification flow makes use of the \emph{xsim} simulator included in \emph{AMD Vivado ML 2023.1}
and executes both the latter and \emph{Spike} on a server featuring
an \emph{Intel Xeon E5-2430} CPU and 64GiB of memory and
running the \emph{Ubuntu 22.04.4 LTS} operating system.
The validation and evaluation of the CPU was carried out by executing
a set of applications from the \emph{Mälardalen WCET} benchmark suite~\cite{Gustafsson_WCET2010},
compiled by using the RISC-V GNU compiler toolchain.
The testbench, as depicted in \figurename~\ref{fig:infrastructure}, instantiates
the CPU under test as well as instruction and data memories.
Verification logic implementing the flow outlined in \figurename~\ref{fig:verification}
manages and coordinates the ISS, through the DPI-C socket, and the various testbench components,
which drive the inputs to the DUT, monitor the DUT and data memory, and
check their changes against the results obtained by the ISS.
A scoreboard collects coverage metrics and statistics.

\begin{table}[t]
	\centering 
	\caption{Number of executed instructions, execution time in terms of clock cycles and microseconds (\textmu s),
	and verification outcome for the execution of \emph{Mälardalen WCET} applications on
	the target superscalar CPU.}
	\begin{tabular}{lrrrr}
		\toprule
		                      &                       & \multicolumn{2}{c}{\textbf{Execution time}} &                      \\ \cmidrule(lr){3-4}
		\textbf{Application}  & \textbf{Instructions} & \textbf{Cycles} &        \textbf{\textmu s} & \textbf{Correctness} \\ \midrule
		\emph{bsort100}       &                  1237 &            5699 &                     74.09 &                   OK \\
		\emph{cnt}            &                  2006 &            7975 &                    103.68 &                   OK \\
		\emph{crc}            &                 21186 &           65796 &                    855.35 &                   OK \\
		\emph{fac}            &                   124 &             538 &                      6.99 &                   OK \\
		\emph{fdct}           &                  1363 &            2438 &                     31.69 &                   OK \\
		\emph{janne\_complex} &                    77 &             487 &                      6.33 &                   OK \\
		\emph{jfdctint}       &                  1748 &            3761 &                     48.89 &                   OK \\
		\emph{lcdnum}         &                   101 &             379 &                      4.93 &                   OK \\
		\emph{matmult}        &                 12156 &           43330 &                    563.29 &                   OK \\
		\emph{prime}          &                  1754 &           14303 &                    185.94 &                   OK \\
		\emph{select}         &                   845 &            2505 &                     32.57 &                   OK \\ \bottomrule
	\end{tabular}
	\label{tab:statistics}
\end{table}

\begin{figure*}[t]
	\centering
	\includegraphics[width=0.98\linewidth]{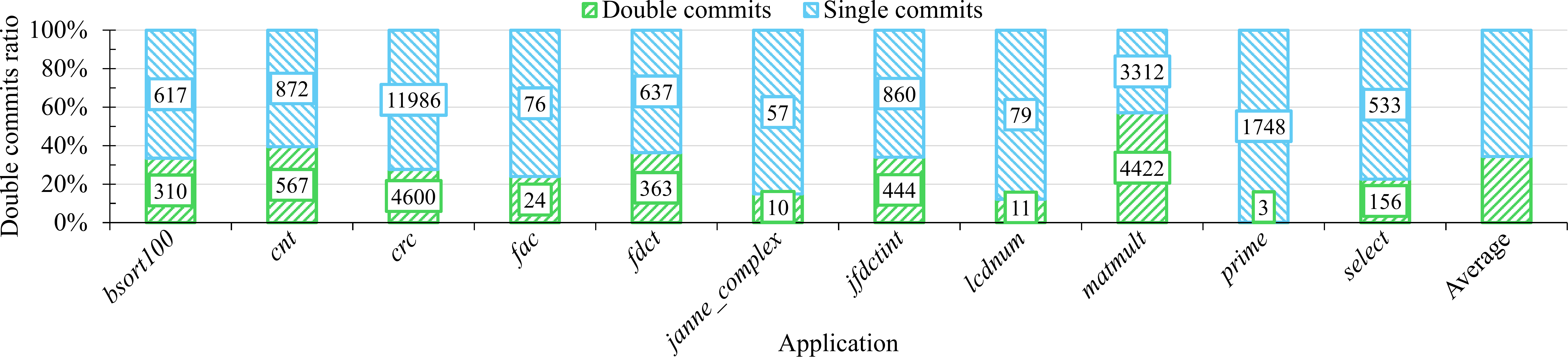}
	\caption{Fraction of double commits when executing the \emph{Mälardalen WCET} applications
	in \tablename~\ref{tab:statistics} on the target superscalar CPU.
	Labels inside the stacked columns refer to the number of double and single commits.}
	\label{fig:commits}
\end{figure*}

\subsection{Experimental results}
\label{ssec:exp_results}
We evaluate the \emph{SupeRFIVe} methodology outlined in Section~\ref{sec:methodology} by
applying it to the dual-issue superscalar CPU described in Section~\ref{ssec:exp_setup} and
executing workloads from the \emph{Mälardalen WCET} benchmark suite
that stress the various parts of the CPU under test by including
integer and floating-point arithmetic, variable-bound loops and loop-iteration-dependent conditions,
nested loops and function calls, and computations on array and matrices.

The functional verification for each of the 11 benchmark applications
shows a correct execution on the CPU under test and delivers an additional set of
performance-related statistics that enable a quick evaluation of how
the superscalar architecture performs with the considered workloads.
For instance, \tablename~\ref{tab:statistics} lists the number of instructions,
execution time in terms of clock cycles and microseconds, and result
of the correctness check, for the sake of brevity, of a select subset of
11 \emph{Mälardalen WCET} benchmark applications.

In particular, the ability to monitor the number of concurrent commits
in the CPU under test provides a measure of the effectiveness of
implementing a superscalar architecture under workloads that show
a significant instruction-level parallelism.
\figurename~\ref{fig:commits} depicts the fraction of double commits
for the execution of each application previously listed in \tablename~\ref{tab:statistics},
with a 34\% average of double commits.
For example, executing \emph{matmult} on the CPU under test results in
4422 double and 3312 single commits, i.e., 57\% of double commits,
over a total of 12156 committed instructions.
The instruction-level parallelism of applications such as \emph{matmult},
which performs the matrix multiplication between
two 20\texttimes20 matrices with nested function calls and triple-nested loops,
is indeed effectively exploitable by the superscalar CPU under test.

\section{Conclusions}
\label{sec:conclusions}
This manuscript introduced the \emph{SupeRFIVe} methodology,
that tackles the problem of performing the functional verification of
superscalar processors in a comprehensive and time-efficient way
by following a simulation-based approach.
In order to do so, it leverages an ISS and
interfaces it through socket communication with
the testbench including the DUT.

In our experimental evaluation, we applied the proposed methodology to
a RISC-V 32-bit RV32IMF dual-issue superscalar core,
using \emph{Spike} as the ISS for RISC-V
and executing \emph{Mälardalen WCET} benchmark applications.
The experiments allowed us verifying the correctness of the processor under test,
and the results listed in manuscript provide an overview of the statistics,
also related to the superscalar nature of the CPU, that
can be collected during its verification.

\bibliographystyle{IEEEtran}
\bibliography{IEEEabrv,2024_icecs}

\end{document}